\documentclass[12pt]{iopart}

\usepackage{iopams}
\usepackage{graphicx}

\begin{document}

\title{Energy spectrum and critical exponents of the free parafermion $Z_N$ spin chain}

\author{Francisco C Alcaraz$^1$, Murray T Batchelor$^{2,3}$ and Zi-Zhong Liu$^{2}$}

\address{$^{1}$Instituto de F\'isica de S\~ao Carlos, Universidade de S\~ao Paulo,
Caixa Postal 369, 13560-970, S\~ao Carlos, SP, Brazil}

\address{$^{2}$Centre for Modern Physics, Chongqing University, Chongqing 400044, China}

\address{$^{3}$Department of Theoretical Physics, 
Research School of Physics and Engineering and Mathematical Sciences Institute, Australian National University, Canberra, ACT 0200, Australia}

\ead{batchelor@cqu.edu.cn}

\begin{abstract}
Results are given for the ground state energy and excitation spectrum of 
a simple $N$-state $Z_N$ spin chain described by free parafermions.
The model is non-Hermitian for $N \ge 3$ with a real ground state energy and a complex excitation spectrum.
Although having a simpler Hamiltonian than the superintegrable chiral Potts model, 
the model is seen to share some properties with it, e.g.,  
the specific heat exponent $\alpha=1-2/N$ and 
the anisotropic correlation length exponents $\nu_\parallel =1$ and $\nu_\perp=2/N$.
\end{abstract}

\vskip 1cm

In 1989 Baxter introduced a simple $Z_N$ Hamiltonian~\cite{Baxter1989a,Baxter1989b}
\begin{equation}
H =  - \sum_{j=1}^L \alpha_j X_j - \sum_{j=1}^{L-1} \gamma_j Z_j Z_{j+1}^{\dagger}
\label{ham0}
\end{equation}
where $\alpha_j$ and $\gamma_j$ are arbitrary parameters.
In matrix form, the operators $X_j$ and $Z_j$ are 
\begin{eqnarray}
X_j &=& I \otimes I \otimes  \cdots \otimes I \otimes X \otimes I \otimes \cdots \otimes I\\
Z_j &=& I \otimes I \otimes  \cdots \otimes I \otimes Z \otimes I \otimes \cdots \otimes I
\end{eqnarray}
where $I$, $X$ and $Z$ are $N \times N$ matrices, with $X$ and $Z$ in position $j$. 
Here $I$ is the identity, with $X$ and $Z$ defined by 
\begin{equation}
X_{jk} = \delta_{j,k+1}, \quad 
Z_{jk} =  \omega^{j-1} \delta_{jk} ,  \quad \hbox{with} \quad \omega = {\mathrm e}^{2\pi {\mathrm i}/N}.
\end{equation}
They satisfy
\begin{equation}
X^N = Z^N = I, \quad X^\dagger = X^{N-1}, \quad Z^\dagger = Z^{N-1}, \quad
Z X = \omega X Z.
\end{equation}

For $N=2$ with $\alpha_j=1$ and $\gamma_j=\lambda$ this Hamiltonian 
is the well known one-dimensional 
quantum Ising model in a transverse field with open 
boundary conditions. 
Based on numerical observations Baxter found that the Hamiltonian (\ref{ham0}) has the 
remarkably simple energy spectrum~\cite{Baxter1989a,Baxter1989b,Baxter2004}
\begin{equation}
-E = \omega^{n_1} \epsilon_1 + \omega^{n_2} \epsilon_2 + \cdots + \omega^{n_L} \epsilon_L
\label{spec}
\end{equation}
for any choice of the integers $n_k = 0, \ldots,  N-1$.
This covers all $N^L$ eigenvalues in the spectrum.
The energy levels $\epsilon_i$ $(i=1,\ldots,L)$ are functions of $\alpha_j, \gamma_j$ $(j=1,\ldots,L)$.
For simplicity we focus our attention here on the choice $\alpha_j=1$ and $\gamma_j=\lambda$, 
with thus 
\begin{equation}
H =  - \sum_{j=1}^L  X_j - \lambda \sum_{j=1}^{L-1} Z_j Z_{j+1}^{\dagger}.
\label{ham}
\end{equation}
The parameter $\lambda$ is a natural generalisation of the Ising transverse field.

Fendley~\cite{Fendley2014} has recently shown that (\ref{spec}) is the energy spectrum of free parafermions.
A key ingredient is the Fradkin-Kadanoff transformation to parafermionic operators 
introduced earlier for the $N$-state clock models~\cite{FK}. 
This is a generalisation of the Jordan-Wigner transformation.
The explicit parafermionic Jordan-Wigner transformation dates back to Morris~\cite{Morris}.
Parafermions have long been considered in both mathematics and physics 
(see, e.g., \cite{Green,Yama,Jaffe2015}).
They have been seen to underpin a range of novel phenomena, particularly with regard to topological phases 
in condensed matter physics~\cite{AF}.

The free parafermionic structure (\ref{spec}) was subsequently confirmed in the related
$\tau_2$ model with open boundaries~\cite{Baxter2014,YP2014,YP2016}, 
from which (\ref{ham0}) and thus (\ref{ham}) follow in the Hamiltonian limit.
The version of the $\tau_2$ model, known as the Bazhanov-Stroganov model~\cite{BS}, 
is a two-dimensional classical chiral spin model connecting 
the six-vertex model and the chiral Potts model in a such a way that the chiral Potts model 
can be viewed as a descendent of the six-vertex model~\cite{BS,Perk2016}. 
The non-Hermitian Hamiltonian (\ref{ham0}) and the related $\tau_2$ model are the 
only known models with an entire spectrum described precisely by free parafermions.

In this Letter we take Fendley's calculations one step further and derive the properties of the energy spectrum for Hamiltonian (\ref{ham}).
To do this, consider the $2L \times 2L$ determinant satisfied by the quasi-particle energies 
$\epsilon_1, \ldots, \epsilon_L$~\cite{Baxter1989a,Baxter1989b,Fendley2014,Baxter2014}, 
\begin{equation}
\left| \begin{array}{ccccccc}
- \epsilon^{N/2} & 1 & 0 & 0 & .. & 0 & 0 \\
1 & - \epsilon^{N/2} & \lambda^{N/2} & 0 & .. & 0 & 0 \\
0 & \lambda^{N/2} & - \epsilon^{N/2} & 1 & .. & 0 & 0 \\
0  & .. & .. & .. & .. & .. & 0 \\
0  & 0 & 0 & 0 & \lambda^{N/2} & - \epsilon^{N/2} & 1 \\
0  & 0 & 0 & 0 & 0 & 1 & - \epsilon^{N/2} \\
\end{array} \right| = 0.
\end{equation}
This is a multinomial in integer powers of $\epsilon^2$ and $\lambda^2$, of degree $L$ in $\epsilon^N$.
This is however, the {\em same} multinomial for all values of $N$ and $\lambda$ if 
one makes the correspondence
\begin{equation}
\epsilon_k^N = \epsilon_k^2(\mathrm{Ising}) \quad {\small \mathrm{and}} \quad \lambda^N = \lambda^2(\mathrm{Ising}).
\end{equation}
The free parafermionic quasi-energies $\epsilon_k$ for the $N$-state model can 
thus be determined from the corresponding solution for the Ising case.  
This latter problem was considered with the necessary open boundary conditions 
long ago by Pfeuty~\cite{Pfeuty}, building on earlier results obtained for the XY model~\cite{LSM1961}.

The solution for arbitrary $N$ and general $\lambda$ is thus given in terms of the quasi-energies 
\begin{eqnarray}
\epsilon_{k_j} &=& \left( 1 + \lambda^N + 2 \lambda^{N/2} \cos {k_j} \right)^{1/N} \cr
&=& \left(1 + \lambda^{N/2} \right)^{2/N} \left( 1 - \theta^2 \sin^2 \frac{{k_j}}{2} \right)^{1/N}, \qquad 
\theta^2 = \frac{4 \lambda^{N/2}}{\left(1 + \lambda^{N/2} \right)^2}, 
\label{theta}
\end{eqnarray}
where the roots $k_j$, $j=1,\ldots,L$, satisfy the equation 
\begin{equation}
{\sin(L+1) k = - \lambda^{N/2} \sin Lk}.
\label{eqn}
\end{equation}
This equation has $L$ roots in the interval $(0,\pi)$ for $\lambda \le (1+1/L)^{2/N}$, while for 
$\lambda > (1+1/L)^{2/N}$ there are $L-1$ roots in $(0,\pi)$, the remaining root being complex.

To obtain the properties of the energy spectrum, first consider the case $\lambda=1$.
The roots of (\ref{eqn}) can then be explicitly written as
\begin{equation}
k_j = \frac{2j\pi}{2L+1}, \quad j=1,\ldots,L.
\label{soln}
\end{equation}
The ground state energy 
\begin{equation}
E_0 = - \sum_{k=1}^L \epsilon_k = - \sum_{k=1}^L  \left(2 \cos \frac{\pi k}{2L+1} \right)^{2/N}
\end{equation}
can be evaluated using the Euler-Maclaurin formula 
\begin{eqnarray}
\sum_{k=1}^L f(k) &=& \int_0^L f(x) dx + B_1 [f(L) - f(0)] \nonumber\\
&& + \sum_{k=1}^\infty \frac{B_{2k}}{(2k)!} \left[ f^{(2k-1)}(L) - f^{(2k-1)}(0) \right]
\end{eqnarray}
where $B_{2k}$ are the Bernoulli numbers. 
In this way we obtain the result
\begin{equation}
E_0(L) = L e_\infty + f_\infty + \frac{\gamma_N}{L^{\nu}} + O\left(\frac{1}{L^{1+\nu}}\right), 
\label{E0}
\end{equation}
with non-integer exponent
\begin{equation}
\nu = \frac2N.
\end{equation}
The bulk and surface energy are
\begin{equation}
e_\infty = - \frac{2^{\nu}}{\sqrt \pi} \frac{\Gamma(\frac12 + \frac{1}{N})}{\Gamma(1 + \frac{1}{N})}, 
\qquad f_\infty =  \frac12 e_\infty + 2^{\nu-1}, 
\end{equation}
where $\Gamma$ is the standard gamma function.
The amplitudes $\gamma_N$ in (\ref{E0}) are given by
\begin{equation}
\gamma_N = - \left[ \frac{1}{N+2} - \sum_{k=1}^{\infty} \frac{2^{2k-1} B_{2k}}{(2k)!} 
\prod_{\ell=0}^{2k-2}\left(\nu-\ell\right) \right] \left(\frac{\pi}{2}\right)^{\nu} 
\end{equation}
with $B_2 = \frac16$, $B_4 = - \frac{1}{30}$, $B_6 = \frac{1}{42},\, \ldots$.
The infinite series only terminates for the case $N=2$, where $\gamma_2 = -\frac{\pi}{24}$.
Note that for $N \ge 3$ the finite-size corrections are no longer governed by integer powers of $1/L$.
This is also the case for the chiral Potts model~\cite{Baxter1989b,ChiralPotts}, 
which although not conformally invariant, does exhibit anisotropic scaling~\cite{Cardy1993}.

\begin{figure}[t]
\begin{center}
\includegraphics[width=1.0\columnwidth]{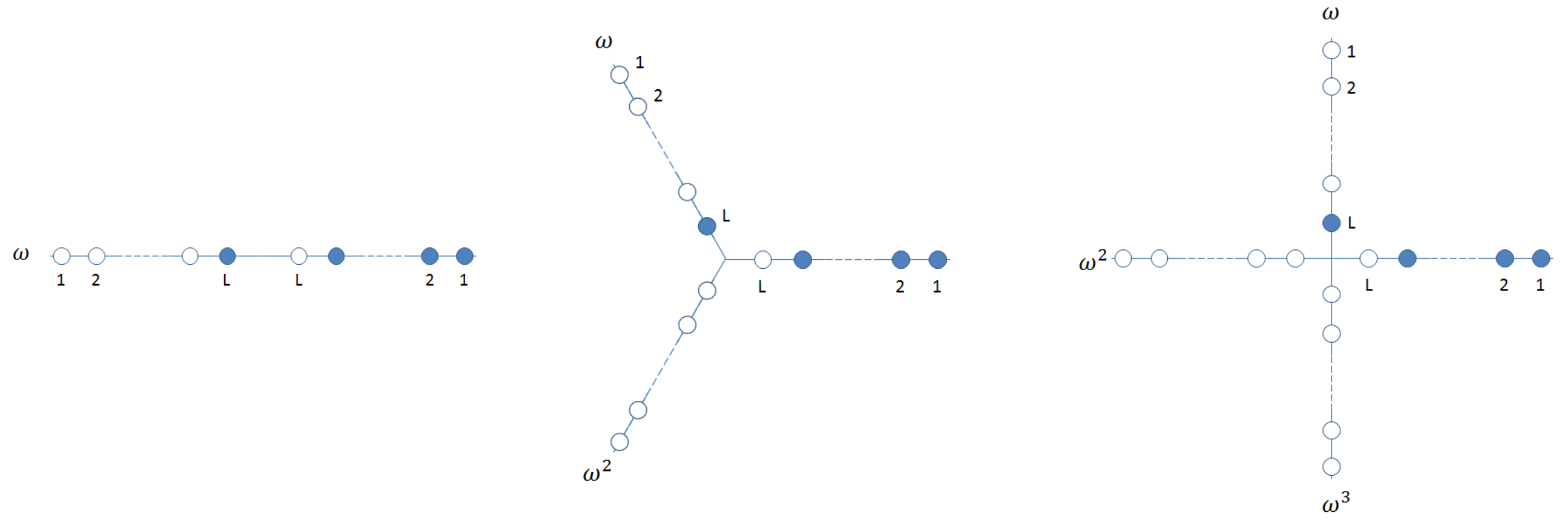}
\caption{$Z_N$ free parafermion quasi-energy levels for 
$N=2~(\omega=-1)$, $N=3~(\omega={\mathrm e}^{2\pi {\mathrm i}/3})$ and 
$N=4~(\omega={\mathrm i})$. In each case a single particle excitation above the groundstate 
is shown on branch $\omega$. Excitations can be on any branch, but 
restricted by the parafermion exclusion rule, effectively defining a Fermi ``exclusion circle" for each level.}
\end{center}
\end{figure}

It is straightforward to derive the excitation spectrum at $\lambda=1$ given $\epsilon_k$ 
and the free parafermion structure of the excitations.
For $N \ge 3$ the Hamiltonian (\ref{ham}) is non-Hermitian, with a complex eigenspectrum. 
The energy spectrum has $Z_N$ symmetry, by which any eigenvalue $E$ has related eigenvalues 
$\omega E, \omega^2 E, \ldots$.
This is the parafermion generalisation of the  $E \leftrightarrow -E$ Ising symmetry. 
Following Fendley~\cite{Fendley2014}, 
the excitation spectrum is best illustrated by diagrams as in figure 1.
As for the Ising $N=2$ case the excitation spectra can be classified according to the number of 
$n$-particle excitations above the Fermi surface. 
For 1-particle excitations we find
\begin{equation}
E(L) - E_0(L) = (2n-1)^\nu \, \varepsilon(p) \left(\frac{\pi}{2L} \right)^{\nu} + O\left(\frac{1}{L^{1+\nu}}\right)
\end{equation}
for $n = 1,\ldots,L$ and $p=1,\ldots, N-1$, with
\begin{equation}
\varepsilon(p) = 1-\omega^p = 1 - \cos\left( {2\pi p}/{N} \right) - \mathrm{i} \sin\left( {2\pi p}/{N} \right).
\end{equation}
Similarly for 2-particle excitations,
\begin{eqnarray}
E(L) - E_0(L) &=& \left[(2m-1)^\nu \, \varepsilon(p) + (2n-1)^\nu \, \varepsilon(q) \right] \left(\frac{\pi}{2L} \right)^{\nu} \nonumber\\
&& \, + O\left(\frac{1}{L^{1+\nu}}\right)
\end{eqnarray}
for $m , n = 1,\ldots,L$ and $p , q =1,\ldots, N-1$ with $m\ne n$.
The generalisation to arbitrary $r$-particle excitations, where $r \le L$ is readily apparent.

For $N=2$ we recover the forms expected from 
conformal invariance~\cite{Cardy1986a,Cardy1986b}, namely 
\begin{eqnarray}
E_0(L) &=& L e_\infty + f_\infty - \frac{\pi \zeta c}{24 L} + O\left(\frac{1}{L^{2}}\right) \\
E_n(L) &=& E_0(L) +  \frac{\pi \zeta (x_n+ r)}{L} + O\left(\frac{1}{L^{2}}\right), \qquad r = 0, 1, 2, \ldots
\end{eqnarray}
with 
\begin{equation}
e_\infty = - \frac{4}{\pi}, \quad f_\infty = 1 - \frac{2}{\pi}, \quad 
\zeta = 2, \quad c = \frac{1}{2}, \quad x_n = n - \frac12.
\end{equation}
The 1-particle excitations are associated with scaling dimension $x_n$, 
the 2-particle excitations with $x_n + x_p, n \ne p$ etc. 
In this way the conformal data of the Ising model, 
$c = \case{1}{2}$, $x_\sigma = x_1 = \case12$, $x_\epsilon = x_1 + x_2 = 2$, 
is recovered for open/free boundary conditions~\cite{BG}.
Unlike for the Ising case, the amplitudes of the finite-size corrections for $N \ge 3$ 
appear to have no physical meaning.

The groundstate energy per site for general $\lambda$ in the $L \to \infty$ limit is  
\begin{equation}
{e_\infty(\lambda) = -\frac{2}{\pi} \left(1 + \lambda^{N/2} \right)^{\nu} 
\int_0^{\pi/2} \left( 1 - \theta^2 \sin^2 x \right)^{1/N} dx} 
\label{einf}
\end{equation}
where $\theta^2$ is defined in (\ref{theta}). 
This result is readily seen to satisfy the duality relation 
$e_\infty(\lambda) = \lambda \, e_\infty(1/\lambda)$. 
The Ising elliptic integral result~\cite{Pfeuty}
\begin{equation}
e_\infty(\lambda) = -\frac{2}{\pi} \left(1 + \lambda \right) E\left(\frac{\pi}{2},\theta \right) 
\end{equation}
is recovered for $N=2$. 
The result (\ref{einf}) can be derived directly using the Euler-Maclaurin formula.
Use can be made of the leading order solutions to equation (\ref{eqn}), which can be 
adapted from the treatment of the XY model with open boundaries~\cite{LSM1961}.
Consider the case $\lambda < 1$. 
Writing 
\begin{equation}
L k_j = \pi j - \pi \kappa_j + O\left( \frac1L \right) , \qquad j = 1, \ldots, L
\end{equation}
gives
\begin{equation}
\cot(\pi \kappa_j) = \frac{\lambda^{N/2} + \cos(\pi j/L)}{\sin(\pi j/L)}
\end{equation}
with solution
\begin{equation}
 \pi \kappa_j  = \frac{\pi j}{2L} + \tan^{-1} \left[ \frac{1-\lambda^{N/2}}{1+\lambda^{N/2}} 
 \tan \left( \frac{\pi j}{2L} \right) \right].
\end{equation}
To leading order, the roots are thus approximated for large $L$ by
\begin{equation}
k_j \approx \frac{\pi j}{L} - \frac{1}{1+\lambda^{N/2}} \frac{ \pi j}{L^2}, \qquad j = 1, \ldots, L.
\end{equation}

On the other hand, for $\lambda > 1$, there is one complex root 
\begin{equation}
k_L = \pi + {\mathrm i} v,
\end{equation}
where $v$ satisfies the equation 
\begin{equation}
\sinh(L+1) v = \lambda^{N/2} \sinh Lv.
\end{equation}
Solving for large $L$, this excitation carries energy 
\begin{equation}
\epsilon_{k_L} = \lambda^{1-L} \left(1 - 2 \lambda^{-N} + \lambda^{-2N} + \ldots  \right)^{1/N}.
\end{equation}
The gap to excitations thus closes exponentially as~${\mathrm e}^{- L \ln \lambda}$ for large $L$.
It follows that the groundstate is $N$-fold degenerate for $\lambda > 1$, 
reflecting the ordered state of the system in this regime.

It is interesting to express $e_\infty(\lambda)$ in terms of hypergeometric functions. 
With the change of variable $x \to \arcsin \sqrt{t}$ in (\ref{einf}) we obtain~\cite{AS}
\begin{equation}
e_\infty(\lambda) = - \left(1 + \lambda^{N/2} \right)^{\nu} F\left( - \case1N, \case12; 1; \theta^2 \right).
\end{equation}
Moreover, using a quadratic relation for the hypergeometric functions~\cite{AS} gives the simple form 
\begin{equation}
e_\infty(\lambda) = - F\left( - \case1N, -\case1N; 1; \lambda^N \right).
\end{equation}
The series representation of $F$ then gives an expansion in $\lambda$, with result 
\begin{equation}
e_\infty(\lambda) = - 1 - \left[ \frac{1}{\pi} \sin\left(\frac{\pi}{N} \right) \Gamma\left( 1 + \frac{1}{N} \right) \right]^2 
\sum_{\ell=1}^{\infty} \frac{\left[\Gamma\left(\ell + \frac{1}{N}\right) \right]^2}{\Gamma(1 + \ell)} \, \frac{\lambda^{N \ell}}{\ell \,!}
\end{equation}
for $\lambda < 1$. The expansion for the case $\lambda > 1$ is obtained from this result using the duality relation 
$e_\infty(\lambda) = \lambda \, e_\infty(1/\lambda)$.

We now turn to critical exponents in the vicinity of the critical point $\lambda=\lambda_c=1$.
For $\lambda < 1$, the mass gap corresponding to 1-particle excitations is 
\begin{equation}
E - E_0 = \varepsilon(p) \left(1-\lambda^{N/2} \right)^{\nu} \quad \mbox{as} \quad L \to \infty 
\label{gap}
\end{equation}
for $p=1,\ldots, N-1$. 
For $N=2$ this is the well known Ising result $E-E_0 = 2(1-\lambda)$~\cite{HB}.
Although in general complex, the mass gap (\ref{gap}) is real when $p=N/2$ for $N$ even.
Analogous results are obtained for the $n$-particle excitations.
The gaps in the excitation spectrum close with a well defined exponent as $\lambda \to 1$.
Writing the mass gaps as $m(\lambda,L)=E-E_0$, 
the dynamical critical exponent $z$ is defined by the scaling behaviour 
$m(\lambda=1,L) \sim L^{-z}$ as $L\to \infty$ and $m(\lambda,L\to\infty) \sim (1-\lambda)^{z}$ as $\lambda \to 1$.
For anisotropic scaling $z=\nu_\perp/\nu_\parallel$, where $\nu_\perp$ and $\nu_\parallel$ 
are the correlation length exponents in the time and space directions, respectively.
From the above results, $z=2/N$ for this model.

Another quantity of interest is the specific heat, defined by
\begin{equation}
C(\lambda,L) = -\frac{\lambda^2}{L} \frac{d^2E_0(\lambda,L)}{d \lambda^2},
\label{specheat}
\end{equation}
which at the critical point $\lambda=1$ scales as $C \sim L^{\alpha/\nu_\parallel}$ as $L \to \infty$.
A lengthy but straightforward calculation gives 
\begin{equation}
C \sim L^{1-2/N} \quad \mbox{as} \quad L \to \infty, 
\end{equation}
implying the result $\alpha/\nu_\parallel = 1 - 2/N$.
Numerical investigation of the specific heat $C(\lambda,L)$ as a function of $\lambda$ and $L$ shows a broad maximum  
at a value $\lambda_c(L)$. 
This peak approaches the critical value $\lambda_c(\infty)=1$, with 
$\lambda_c(L) - \lambda_c(\infty) \sim L^{-\nu_\parallel}$ as $L \to \infty$. 
%
%\begin{equation}
%\lambda_c(L) - \lambda_c(\infty) \sim L^{-\nu_\parallel}, \quad \mbox{as} \quad L \to \infty,
%\end{equation}
%
We have confirmed this scaling relation numerically, 
with exponent value $\nu_\parallel =1$ independent of $N$.

Collecting the results, the free parafermion $Z_N$ spin chain has critical exponents 
\begin{equation}
\alpha = 1 - 2/N, \quad \nu_\parallel =1, \quad \nu_\perp = 2/N.
\end{equation}
These are the same values derived for the 
superintegrable chiral Potts model~\cite{Baxter1989b,ChiralPotts}. 
The specific heat exponent is also that of the $Z_N$ Fateev-Zamolodchikov model~\cite{FZ}.

We conclude with some remarks.
In general non-Hermitian Hamiltonians describe the dynamics of physical systems that are not conservative.
The quantum Hamiltonian of the chiral Potts model is Hermitian, with a real eigenspectrum. 
Hamiltonian (\ref{ham}) is an example from the 
class of models which are non-Hermitian, with a complex eigenspectrum for $N\ge 3$. 
Nevertheless, the model has a real ground state and a remarkably simple excitation spectrum governed 
by the structure of free parafermions.
We have seen here that the eigenspectrum shares some properties with the chiral Potts model. 
As remarked by Cardy~\cite{Cardy1993} in discussion of the chiral Potts model 
from the perspective of conformal field theory, 
several of the usual properties of Hermitian systems, 
such as insensitivity of bulk thermodynamic quantities to boundary conditions, 
can fail in the non-Hermitian case.
This note of caution should apply even more so for the model under consideration here.  
Given this point, along with the burgeoning  relevance and interest in the physics of parafermions, 
this is clearly a model deserving of further attention.

\ack 

We thank Jacques Perk for instructive comments on the first version of our article.
This work benefited from participation by FCA and MTB 
at the Isaac Newton Institute for Mathematical Sciences programme ``Mathematical Aspects of Quantum Integrable Models in and out of Equilibrium" in Cambridge during January 2016. 
The work of FCA was supported in part by the Brazilian agencies FAPESP and CNPq. 
MTB acknowledges support from the 1000 Talent Plan and Chongqing University in China. 
The work of MTB has also been supported by National Natural Science Foundation of China Grant No.~11574405 and Australian Research Council Discovery Project DP130102839.

\end{document}